\documentclass[%
twocolumn,
amsmath,amssymb,
aps,
pra,
]{revtex4-2}

\usepackage{graphicx}
\usepackage{dcolumn}
\usepackage{bm}
\usepackage{hyperref, color}

\begin{document}
	
	\preprint{APS/123-QED}

\title{Perfect stimulated Raman adiabatic passage with imperfect finite-time pulses}
	
	\author{Shruti Dogra}
	\email{shruti.dogra@aalto.fi}
	
	\author{Gheorghe Sorin Paraoanu}
	\email{sorin.paraoanu@aalto.fi}
	\affiliation{QTF Center of Excellence, Department of Applied Physics, Aalto University School of Science, P.O. Box 15100, FI-00076 AALTO, Finland}

	\date{\today}%
	
	\begin{abstract}
We present a well-tailored sequence of two Gaussian-pulsed drives that achieves  perfect population transfer in STImulated Raman Adiabatic Passage (STIRAP). We give a theoretical analysis of the optimal truncation and relative placement of the Stokes and pump pulses.
Further, we obtain the power and the duration of the protocol for a given pulse width. Importantly, the duration of the protocol required to attain a desired value of fidelity depends only logarithmically on the infidelity.
Subject to optimal truncation of the drives and with reference to the point of fastest transfer, we obtain a new  adiabaticity criteria, which is remarkably simple and effective.
	\end{abstract}

	\maketitle

	\section{Introduction}
	The STIRAP (Stimulated Raman Adiabatic Passage) protocol got its first validation in an experiment ~\cite{gaubatz-1988, gaubatz-1990}
	where partially overlapping Stokes and pump laser beams were employed to transfer the population 
	from a lower energy state to a higher vibrational state without 
	populating the intermediate level in a three-level system consisting of molecular vibrational states. 
	This was done with a non-trivial pulse arrangement (usually referred to as counter-intuitive sequence), 
	where the  Stokes pulse precedes the pump pulse.
	This selective and precise adiabatic transfer of
	population has been a subject of much interest from  a theoretical as well as experimental perspective~\cite{shore-1995, Shore-2017}.
	Due to its intrinsic robustness against practical imperfections, STIRAP has been widely adopted 
	in various different experimental systems~\cite{shore-review-2017, Bergmann_2019}.
    Most importantly, the success of the protocol (even from the theoretical point of view) relies on the fulfilment of the adiabaticity criteria~\cite{kulinski-1989,molmer-pra, daniel-pra-2009, tong-prl-2007, amin-prl-2009, tong-prl-2005}. 
    As per quantum adiabatic theorem~\cite{Born1928, Schiff_QM}, the system, which is initialised in an eigenstate, follows the corresponding eigenstate of the instantaneous Hamiltonian.
	However, a widely acceptable quantitative criteria for adiabaticity is still lacking \cite{tong-prl-2010}. 
	An interesting approach based on local adiabaticity criteria is discussed in~\cite{roland-pra-2002}, where the Hamiltonian generating adiabatic evolution is designed in a such a way that it fulfils the local adiabaticity condition at infinitesimal time steps, which is further used to obtain the adiabatic-evolution version of the Grover's search algorithm.

	Here we present an adiabaticity criteria which is demonstrably  sufficient for achieving perfect population transfer. Our criteria is markedly different from the existing ones and is surprisingly effective despite its simplicity.
	The key concept of our analysis is based on the most sensitive point of the dynamics, which is in the middle of the sequence, where the rate of evolution of the quantum state is the highest. Also, a high-fidelity STIRAP requires the pulse sequence to be implemented in an optimal time, which involves optimal truncation of the drives, as well as optimal width of the drives and relative placement of the drives in the pulse sequence. Here we show how these issues can be solved.
	Another important aspect is the power of the pulses, which we obtain optimally with the help of our newly introduced adiabaticity criteria. 
	We analyse the situation in detail and arrive at analytical expressions that lead to a perfectly tailored STIRAP. 
	
	The protocol studied here is experimentally implementable to any three-level system. The set of parameters presented here can be directly used in a circuit QED based experimental setup with a multi-level Josephson-junction artificial atom~\cite{kumar-nature-2016}.  
	There are also ways to suppress the non-adiabatic excitations by employing shortcuts to adiabaticity~\cite{Bergmann_2019}. In superadiabatic(sa)-STIRAP in three-level systems, an additional counterdiabatic pulse is needed, realizing direct coupling between the initial and the target states.  A circuit QED based setup implementing saSTIRAP protocol in a three-level system has been demonstrated in Ref.~\cite{antti-science-2019} and its robustness against various experimental imperfections has been analysed in Ref.~\cite{sd-arxiv-2022}.

	\par 
The Hamiltonian governing the STIRAP for a three-level system in the computational basis 
	$\{ \vert 0 \rangle$, $\vert 1 \rangle$, $\vert 2 \rangle\}$, in the dispersive regime and under the rotating wave approximation,
	is given by \cite{kumar-nature-2016}
	\begin{equation}
	  H_0 = \frac{\hbar}{2}\left( \begin{array}{ccc}
	 0 & \Omega_{01}(t) & 0 \\
	 \Omega_{01}(t) & 2\delta_{01} & \Omega_{12}(t) \\
	 0 & \Omega_{12}(t) & 2(\delta_{01}+\delta_{12})
	 \end{array}  \right),
	 \end{equation}
	where the time-varying amplitudes of the driving fields are chosen as Gaussians 
	with equal 
	standard deviation $\sigma$.  These Gaussians are separated in time by an amount $t_s$ given by
	\[ \Omega_{01}(t) = \Omega_{01}^{0} e^{-t^2/2\sigma^2} \quad \textrm{and} \quad  \Omega_{12}(t) = \Omega_{12}^{0} e^{-(t-t_s)^2/2\sigma^2}.\] 
	
	An adiabatic evolution is ideally infinitely slow and would require the system to be in 
	an eigenstate of the instantaneous Hamiltonian at all times. At the two-photon resonance condition ({\it i.e.} $\delta_{01}=-\delta_{12}$),
	a convenient choice of the eigenvector is the dark state $\vert {\rm D} \rangle=\cos \Theta \vert 0 \rangle - \sin \Theta \vert 2 \rangle$, 
	which does not have any dependence on the intermediate level $\vert 1 \rangle$. Here the mixing angle $\Theta$ is defined by 
	$\Theta = \tan^{-1}\Omega_{01}(t)/\Omega_{12}(t)$. 

\section{Optimal pulse duration}
Adiabatic drive in principle demands infinitely long operation time for a complete 
transfer of population. Ideally, as required by STIRAP, Gaussians pulses are of infinite extent.
However, to cope with the experimental limitations on pulse generation and to minimize the losses
due to decoherence, one would have to truncate the Gaussians $\Omega_{01}(t)$ and $\Omega_{12}(t)$ optimally. Therefore there is a tradeoff between the loss in the transfer fidelity that can be afforded and the total pulse time. Revisiting the mixing angle, while assuming $\Omega_{01}^{0}=\Omega_{12}^{0},$ we write
\begin{equation}
 \tan \Theta(t) = \frac{\Omega_{01}(t)}{\Omega_{12}(t)}=\frac{e^{-rt/\sigma}}{e^{-r^2/2}}, \label{eq:theta_r}
\end{equation}
where we introduce the parameter $r=t_s/\sigma$. 

\begin{figure}
 \centering
 \includegraphics[scale=1,keepaspectratio=true]{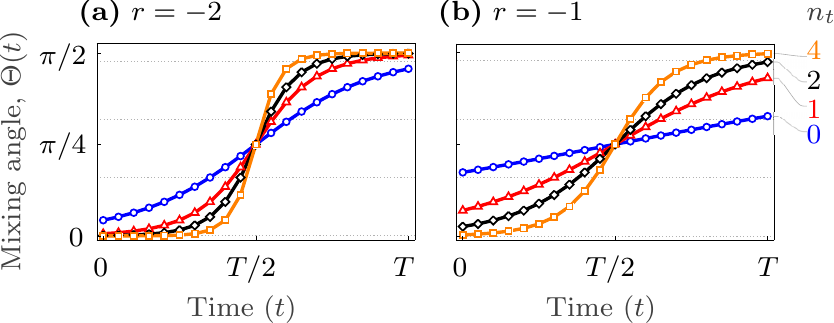}
 \caption{Time variation of $\Theta(t)$ versus total pulse duration $T$ for different values of $n_t$
 is shown at two different values of $r$: (a) $r=-2$ and (b) $r=-1$.
 In each of these figures, blue curve with circular markers, red curve with triangles, black curve
 with diamonds, and orange curve with squares correspond to $n_t=0,1,2,4$ respectively (also specified
 at the right end of each curve in part (b)). \label{fig-vary-n}}
\end{figure}

\begin{figure}
 \centering
 \includegraphics[scale=1,keepaspectratio=true]{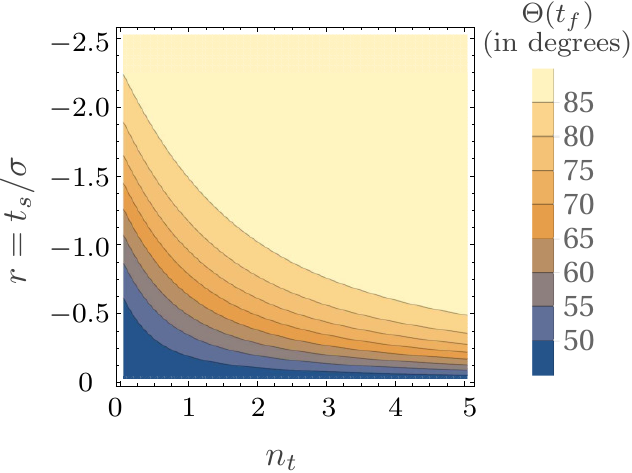}
 \caption{Contour plot showing the mixing angle at time $t=t_f$ corresponding to the normalized relative separation $r$ between the two 
 driving fields and the parameter $n_t$ characterizing the transfer time. \label{fig-contour-thetaf}}
\end{figure}

We truncate this STIRAP pulse sequence (consisting of drives $\Omega_{01}(t)$ and $\Omega_{12}(t)$) from left at time $t=t_i= -n_t \sigma + t_s = - (n_t-r)\sigma$, which we call initial time point,  and from right at $t=t_f=n_t  \sigma$, which we call final time point, where $n_t$ is a real number ($n_t \in \mathbb{R}$). 
The total pulse duration is therefore $T=t_f-t_i=(2 n_t - r) \sigma$.
We fix the values of $r$ and $\sigma$ and present the corresponding dynamics of $\Theta(t)$ versus total pulse duration as shown in
Fig.~\ref{fig-vary-n}, where  different curves correspond to different values of $n_t$.
The width of the Gaussian may be fixed to any arbitrary value (here $\sigma=30$ ns) as this does not effect the variation of $\Theta(t)$ in a given total time $T$.
Ideally, during the STIRAP drive, the mixing angle $\Theta(t)$ is expected to vary from $0$ to $\pi/2$, while in reality, a finite-time sequence effectively varies $\Theta(t)$ from $\Theta_i \rightarrow 0$ 
to $\Theta_f=\pi/2-\Theta_i$. A closer look at Fig.~\ref{fig-vary-n} immediately concludes that a choice
of small enough $n_t$ might result in a large $\Theta_i$ \emph{e.g.} in Fig.~\ref{fig-vary-n}(b) the  blue curve marked with circles has $n_t=0$ and $\Theta_i>\pi/6$ while the orange curve marked with squares corresponds to $n_t=4$ and $\Theta_i \approx 0$. Thus, for a given $r$, lower values of $n_t$ result in a poor transfer fidelity.
To make it worse, real situations by default have $\Theta_i \approx 0$, therefore a too small value of $n_t$ is susceptible to create errors which can be difficult to trace. 
An elaborated picture of the ideal situation is presented in Fig.~\ref{fig-vary-n}(b), where curves corresponding to $n_t=1,4$ have $\Theta_i = 12.6^0$ and $0.64^0$ respectively, with total pulse duration being $105$ ns and $315$ ns respectively. 
Clearly in this case $n_t=4$ has a much more desirable outcome than that of $n_t=1$, despite the high time cost.
Another important factor that plays a role in the time management of the STIRAP implementation is the relative separation between the two pulses ($r=t_s/\sigma$). Comparing curves corresponding to
 $n_t=1$ in Fig.~\ref{fig-vary-n}(a) and (b), it is found that for $r=-2$, $\Theta_i \approx 1^0$
 and the total pulse duration is $140$ ns. Thus an optimal combination of $n_t$ and $r$ provides an
 efficient STIRAP without compromising much with respect to the time cost. A quite thorough
 picture can be obtained from the contour plot in Fig.~\ref{fig-contour-thetaf}, wherein 
 mixing angle corresponding to the final state ($\Theta_f=\pi/2-\Theta_i$) is plotted for different combinations of $n_t$ and $r$.
 The final value of the mixing angle ranges from $\Theta_f=45^0$ (corresponding to $r=0, n_t=0$ and thus no time evolution)
 to $\Theta_f=90^0$, which corresponds to a complete transfer of population from $\vert 0 \rangle \rightarrow \vert 2 \rangle$.

\par
 Next, we quantify the threshold for a desired fidelity of the STIRAP protocol. From Eq.~(\ref{eq:theta_r}),
\begin{eqnarray}
 -\frac{rt}{\sigma} &=& -\frac{r^2}{2} + \ln \left[ \tan \Theta(t) \right]. \label{eq-theta2}
\end{eqnarray} 
\par 
As stated earlier, due to the STIRAP driven dynamics, our three-level system is in the dark state $| {\rm D} \rangle$ at all times.
We parametrize the initial dark state with close proximity to the ground state $\vert 0 \rangle$ by

\begin{equation}
  \vert \psi_i \rangle = 
 \left( \begin{array}{c}  \cos \Theta_i \\ 0 \\ -\sin \Theta_i   \end{array} \right) 
                        = \left( \begin{array}{c} \sqrt{1-\epsilon^2} \\ 0 \\ -\epsilon \end{array} \right), \label{eq-si}
 \end{equation}
such that the ideal case is recovered when  $\epsilon \longrightarrow 0$. Since we know that during the 
evolution under the STIRAP Hamiltonian our system is in the dark state at all times, the final state reads

\begin{equation}
  \vert \psi_f \rangle  =  \left( \begin{array}{c}  \cos \Theta_f \\ 0 \\ -\sin \Theta_f   \end{array} \right)= \left( \begin{array}{c} \epsilon  \\ 0 \\ -\sqrt{1-\epsilon^2} \end{array} \right). \label{eq-sf}
 \end{equation}
 For both the initial and final states, the parameter $\epsilon$ is a measure of infidelity. Indeed, the fidelity is $F_{i} = F_{f} = |\langle 0|\psi_{i}\rangle |^2 = |\langle 2|\psi_{f}\rangle |^2 = 1-\epsilon^2$ and therefore the infidelity is $\epsilon^2$. At the final time point, $t=t_f=n_t \sigma$, and from Eqs.~(\ref{eq-theta2}),~(\ref{eq-si}), and~(\ref{eq-sf}), 
 one may easily arrive at
 \begin{equation}
  n_t = \frac{1}{r} \ln \frac{\epsilon}{\sqrt{1-\epsilon^2}} +\frac{r}{2} \label{eq-nt}.
 \end{equation} 
 Thus the total pulse duration is
   \begin{equation}
   T =  \frac{2\sigma}{r} \ln \frac{\epsilon}{\sqrt{1-\epsilon^2}}, \label{eq-T} 
  \end{equation} 
  which, as expected, is directly proportional to the widths of the Gaussians. A larger 
  value of $|r|$ corresponds to faster truncation (smaller $n_t$) and is overall advantageous 
  in terms of the total pulse duration.
  For small enough $\epsilon$ ($\epsilon^2 << 1$), Eq.~\ref{eq-T} leads to $\epsilon \propto e^{rT/(2\sigma)}$. Thus, the infidelity decreases 
  exponentially with total time.

\begin{figure}
 \centering
 \includegraphics[scale=1,keepaspectratio=true]{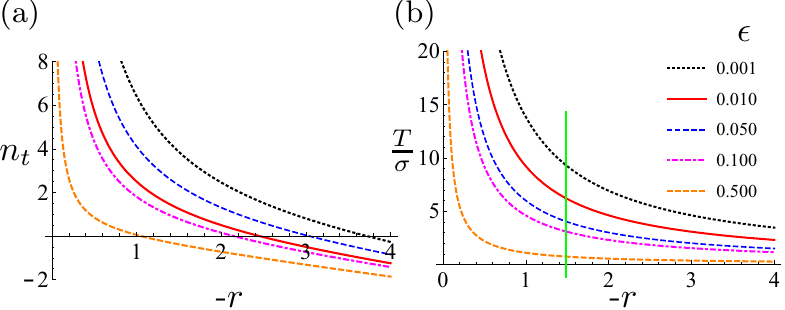}
 \caption{Plots of (a) $n_t$ vs $-r$ and (b) $T/\sigma$ vs $-r$ corresponding to different values of $\epsilon$.}
 \label{fig_r_vs_nt}
\end{figure}

  Fig.~\ref{fig_r_vs_nt}(a) contains plots of $n_t$ vs $-r$, where different curves correspond to 
  different values of $\epsilon$. It is interesting to note that there may exist the STIRAP sequence  
  even for negative values of $n_t$
   when the relative separation between the two Gaussians is large.
  However, the total transfer time is positive as expected, which is clearly seen in Fig.~\ref{fig_r_vs_nt}(b) 
  showing the variation of $T/\sigma$ with $-r$ at corresponding values of $\epsilon$.
  Consider the vertical green line at $r=-1.5$ in Fig.~\ref{fig_r_vs_nt}(b), and the values 
  of $T/\sigma$ while it intersects different curves plotted at different values of $\epsilon$.
  The smaller $\epsilon$ is, the higher the fidelity, which requires larger values of $T/\sigma$ for 
  a fixed value of $r$. For instance, assuming $\sigma=40$ ns, a total transfer time of $184.2$ ns, $122.8$ ns,
  $79.9$ ns, $61.3$ ns, and $14.6$ ns are required to obtain final values of the mixing angle $\Theta_f$ to be 
  $89.9^0$, $89.2^0$, $85.9^0$, $81.9^0$, and $48.6^0$ respectively.

Another interesting situation arises when $\Omega_{01}^{0} \neq \Omega_{12}^{0}$, which influences the left and right truncation limits, 
however the total transfer time remains unchanged. This situation is discussed in detail in Appendix~\ref{app-omega01neqomega12}.

\section{Adiabaticity criteria}

Next, we evaluate the optimal value of the pulse amplitude corresponding to the optimal transfer time calculated in the last section.
The total pulse area then may be compared with the total energy required to achieve the selective population transfer. 
The adiabatic criteria for a STIRAP implementation implies that at any 
	arbitrary time $t$ the effective area $A (t)$ is much greater than the 
	time rate of change of the mixing angle,
	\begin{equation}
	A(t)=  \sqrt{\Omega_{01}^2(t)+\Omega_{12}^2(t)} \quad >> \quad \dot{\Theta}(t), \label{adia1}
	\end{equation}
which upon integration, gives rise to the global adiabaticity condition, as discussed in~\cite{kumar-nature-2016}.

\begin{figure}[ht]
 \centering
 \includegraphics[scale=0.92,keepaspectratio=true]{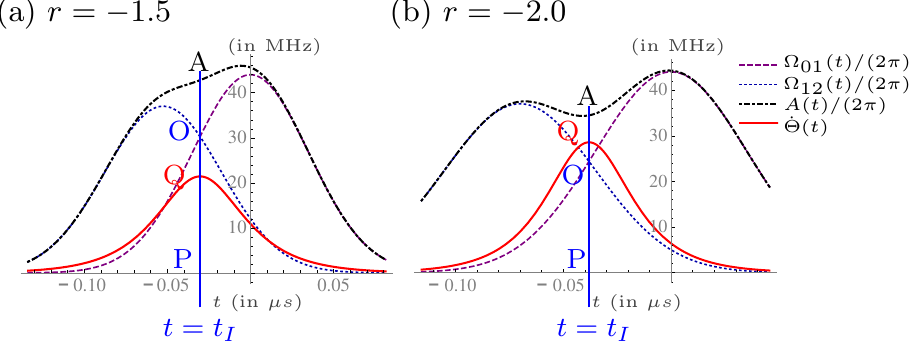}
 \caption{Plots of the time-varying amplitudes $\Omega_{01}(t)/(2\pi)$ and $\Omega_{12}(t)/(2\pi)$, effective area $A(t)/(2\pi)$,
and $\dot{\Theta}(t)$ versus time are shown for (a) $r=-1.5$ and (b) $r=-2$. Here  $\sigma=35$ns,
$\Omega_{01}^{0}/(2\pi)=44$ MHz, and $\Omega_{12}^{0}/(2\pi)=37$ MHz.}
 \label{fig:adia}
\end{figure}

For a deeper insight into the protocol, let us look at the time dependence of these  quantities.
The plots of the time-varying amplitudes of the driving fields ($\Omega_{01}(t)$ and $\Omega_{12}(t)$), effective area $A(t)$,
and rate of change of the mixing angle $\dot{\Theta}(t)$ are shown in Fig.~\ref{fig:adia}, where $\sigma=35$ ns,
$\Omega_{01}^{0}/(2\pi )=44$ MHz, and $\Omega_{12}^{0}/(2\pi )=37$ MHz. 
It is easy to notice that the rate of change of mixing angle is maximum in the middle of the sequence where $\Omega_{01}(t) = \Omega_{12}(t)$.
 
Ideally (for $\epsilon \rightarrow 0$), at $t=t_I$, the mixing angle is $\Theta=\pi/4$ and the populations are $p_0=p_2=0.5$. Also the STIRAP sequence is fastest and more prone to errors at this point, such that a non-zero $p_1$ occurs close to this point.
Thus $t=t_I$ plays an important role. It is also clearly seen in Fig.~\ref{fig:adia} that the variation of the mixing angle with time attains its maximum value at $t=t_I$, when $\Omega_{01}(t) = \Omega_{12}(t)$.

Thus, in most of the cases, especially the ones corresponding to poor performance of STIRAP, the time $t=t_I$ corresponds to close values of the terms on the left and right hand sides of
the inequality in Eq.~(\ref{adia1}). In Fig.~\ref{fig:adia}, at $t=t_I$, a vertical blue line intersects 
various curves, such that $A(t_I)/(2\pi)$, $\Omega_{01}(t_I)/(2\pi)=\Omega_{12}(t_I)/(2\pi)$, $\dot{\Theta}(t_I)$ and the time axis are
labelled by points A, O, Q, and P respectively. An intuitive argument based on observation leads to a non-trivial 
relation, that has to be obeyed for a better performance of the STIRAP, given by,
\begin{equation}
	\Omega_{01}(t_I) - 2\pi\dot{\Theta}(t_I) \quad \gtrsim \quad \sqrt{\Omega_{01}^2(t_I)+\Omega_{12}^2(t_I)} - \Omega_{01}(t_I), \label{adia2}
	\end{equation}
or	alternatively,
	\begin{equation}
	\frac{ \Omega_{01}(t_I) - 2\pi\dot{\Theta}(t_I)}{\Omega_{01}(t_I)} \quad \gtrsim \quad (\sqrt{2}-1), \label{adia3}
	\end{equation}
	as $\Omega_{01}(t)=\Omega_{12}(t)$ at $t=t_I$. Time ($t_I$) is given by,
	\begin{equation}
	 t_I=\frac{t_s}{2}+ \frac{\sigma^2}{t_s} \ln \alpha, \label{adia4}
	\end{equation}
where $\alpha=\Omega_{01}^{0}/\Omega_{12}^{0}$. Further,
\begin{eqnarray}
 \dot{\Theta}(t) &=& \frac{\dot{\Omega}_{01}(t) \Omega_{12}(t)-\Omega_{01}(t) \dot{\Omega}_{12}(t)}{\Omega_{01}^2(t)+\Omega_{12}^2(t)}  \nonumber \\
 \end{eqnarray}
 and at
 $t=t_I$, 
 
 \begin{eqnarray}
 \quad \dot{\Theta}(t_I) &=& -\frac{t_s}{2\sigma^2}. \label{adia5}
\end{eqnarray}

\begin{figure}[ht]
 \centering
  \includegraphics[width=8cm,keepaspectratio=true]{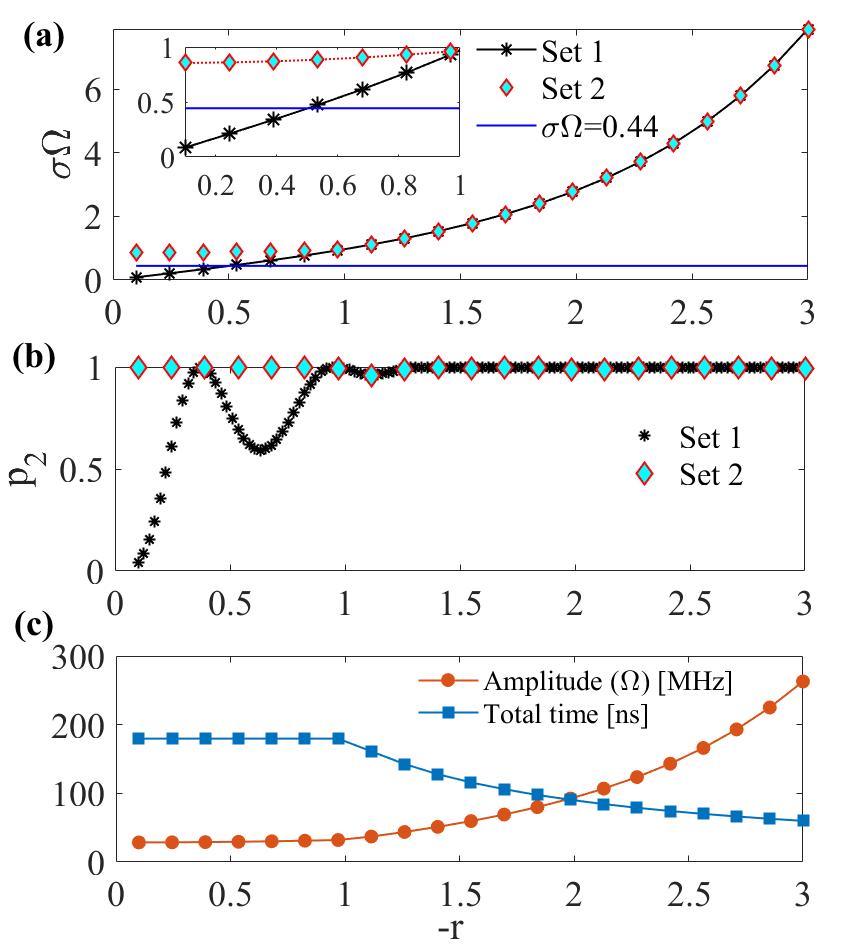}
 \caption{Plots for (a) STIRAP adiabaticity criteria ($\sigma\Omega=R.H.S.$) and (b) corresponding final population in the second excited state ($p_2$) for Set 1 (Eqs.~(\ref{eq-nt}) and~(\ref{adia7})) with continuous black curve with markers, for Set 2 (Eqs.~\ref{eq-nt-r} and~\ref{adia7r}) diamond markers and the global adiabaticity condition with continuous blue line are shown. The inset shows an elaborated view of the respective quantities for $|r|<1$. 
 (c) Total pulse duration in blue with square markers and drive amplitude in red with circular markers are shown for Set 2, where $\epsilon=0.05$, $\alpha=1$, and $\sigma=30$ ns.}
 \label{fig:sigmaomega}
\end{figure}

Note that even for $\alpha \neq 1$, $\dot{\Theta(t)}$, attains its maximum value at $t=t_I$, where $\Omega_{01}(t)=\Omega_{12}(t)$.
 Irrespective of the asymmetry introduced by different pulse amplitudes ($\Omega_{01}^{0}$ and $\Omega_{12}^{0}$), the rate of population transfer reaches its maximum at $t=t_I$. 
From Eqs.~\ref{adia2}-- \ref{adia5}, the condition for a better STIRAP result is given by,
\begin{equation}
 \sigma \frac{\Omega_{01}^{0}}{2\pi} \gtrsim \frac{-r e^{t_I^2/2\sigma^2}}{2(2-\sqrt{2})}. \label{adia6}
\end{equation}
For $\Omega_{01}^{0}/(2\pi)=\Omega_{12}^{0}/(2\pi)=\Omega$, $\alpha=1$, and the above inequality results into
\begin{equation}
 \sigma \Omega \gtrsim \frac{-r e^{r^2/8}}{2(2-\sqrt{2})}. \label{adia7}
\end{equation}
This is the adiabaticity criteria for the STIRAP population transfer, which is obtained by assuming the system to be in the dark state at all times (see Eq.~\ref{eq-si}).
The STIRAP population transfer calculated using Eqs.~\ref{eq-nt} and~\ref{adia7} (for $\alpha=1$) will be  labelled in the following as parameter `Set 1'. The dependence of the right side of Eq.~\ref{adia7}  with respect to $-r$ is plotted as shown in Fig.~\ref{fig:sigmaomega}(a) with continuous black curve with markers. The corresponding population transfer obtained from Set 1 (for $\epsilon=0.05$) is shown in Fig.~\ref{fig:sigmaomega}(b) with black markers. It is noteworthy that the plot of $p_2$ versus $-r$ (see Fig.\ref{fig:sigmaomega}(b)) is independent of the values of $\sigma$ and is in fact dependent on $\epsilon$. On the other hand, the total pulse duration T is directly proportional to $\sigma$ and the drive amplitude $\Omega$ is inversely proportional to $\sigma$. 
Furthermore, a larger value of $\epsilon$ leads to less efficient population transfer in significantly shorter time.

The results of Eq.~(\ref{adia7}) are compared with the global adiabaticity criteria, $\sigma \Omega >> \sqrt{\pi}/4$ \cite{kumar-nature-2016}.
   In Fig~\ref{fig:sigmaomega}(a), the continuous blue line corresponds to $\sigma \Omega = \sqrt{\pi}/4$, which means that for the global adiabaticity criteria to be satisfied, the product $\sigma \Omega$ must lie significantly above the blue line. 
On comparison between the continuous blue curve and black curve with markers in Fig~\ref{fig:sigmaomega}(a), we find that it is possible to have an effective population transfer even when the global adiabaticity criteria is clearly violated in the region $|r|<0.5$. For instance, considering a STIRAP evolution with $r=-0.4$, $\sigma=5$ ns, $\Omega=70$ MHz, and optimally tailored time of $115$ ns (where $n_i=n_f=11.3$ with $\epsilon=0.01$), starting from the ground state yields a final state population, $p_2\approx0.99$ that goes beyond 
the global adiabaticity condition (stated earlier), and $T \Omega = 8.02$, which is $<10$ and hence violates the adiabaticity condition reported in~\cite{shore_rmp_1998, sorin_2017}. Also, for larger values of $|r|$, Fig~\ref{fig:sigmaomega}(a) presents much disparity between the two adiabaticity criteria.

Further, on close observation of the $p_2$ population profile in Fig~\ref{fig:sigmaomega}(b), we find that the population transfer is not very efficient for $|r|<1$, and that the above example of a perfect transfer at $r=-0.4$ is a mere coincidence. These imperfections originate in the assumption of dark state dynamics, which is not valid for small values of $|r|$ due to spurious excitations.
 They can be compensated by higher power of the drive.
 Based on these phenomenological considerations, we can design  an optimal set of parameters for $|r|<1$. We call this $r-$conditioned set of parameters and label it as Set 2, given by 

\begin{eqnarray}
  n_t &=& -\ln \frac{\epsilon}{\sqrt{1-\epsilon^2}} +\frac{r}{2}, \quad |r|<1 \nonumber \\
   &=& \frac{1}{r} \ln \frac{\epsilon}{\sqrt{1-\epsilon^2}} +\frac{r}{2}, \quad |r| \geq 1    \label{eq-nt-r} \\
  \sigma \Omega &\gtrsim& \frac{e^{r^2/8}}{2(2-\sqrt{2})}, \quad |r|<1 \nonumber \\
  &\gtrsim& \frac{-r e^{r^2/8}}{2(2-\sqrt{2})}, \quad |r| \geq 1, \label{adia7r}
\end{eqnarray}
where $\alpha=1$. 
The plot for the right hand side of the inequality in Eq.~(\ref{adia7r}) versus $r$ is shown as diamond markers in Fig~\ref{fig:sigmaomega}(a) and the corresponding final population $p_2$, calculated from Set 2 is shown in ~\ref{fig:sigmaomega}(b). Clearly, Fig~\ref{fig:sigmaomega}(b) reflects a perfect STIRAP transfer for parameter Set 2. Fig.~\ref{fig:sigmaomega}(c) shows the variation of total pulse duration (T) and the drive amplitude ($\Omega$) as functions of $r$, where $\sigma=30$ ns.

We also simulated STIRAP where the product $\sigma \Omega$ is close to but lesser than the respective right hand sides of Eq.~(\ref{adia7r}), and find that we can still arrive at a good enough population transfer in certain situations. This is especially true for large values of $|r|$. Thus, we conclude that the adiabaticity condition in Eq.~(\ref{adia7r}) is sufficient but not necessary for a perfect population transfer.

\section{A perfect STIRAP protocol}

\begin{figure}[ht]
 \centering
 \includegraphics[width=8cm,keepaspectratio=true]{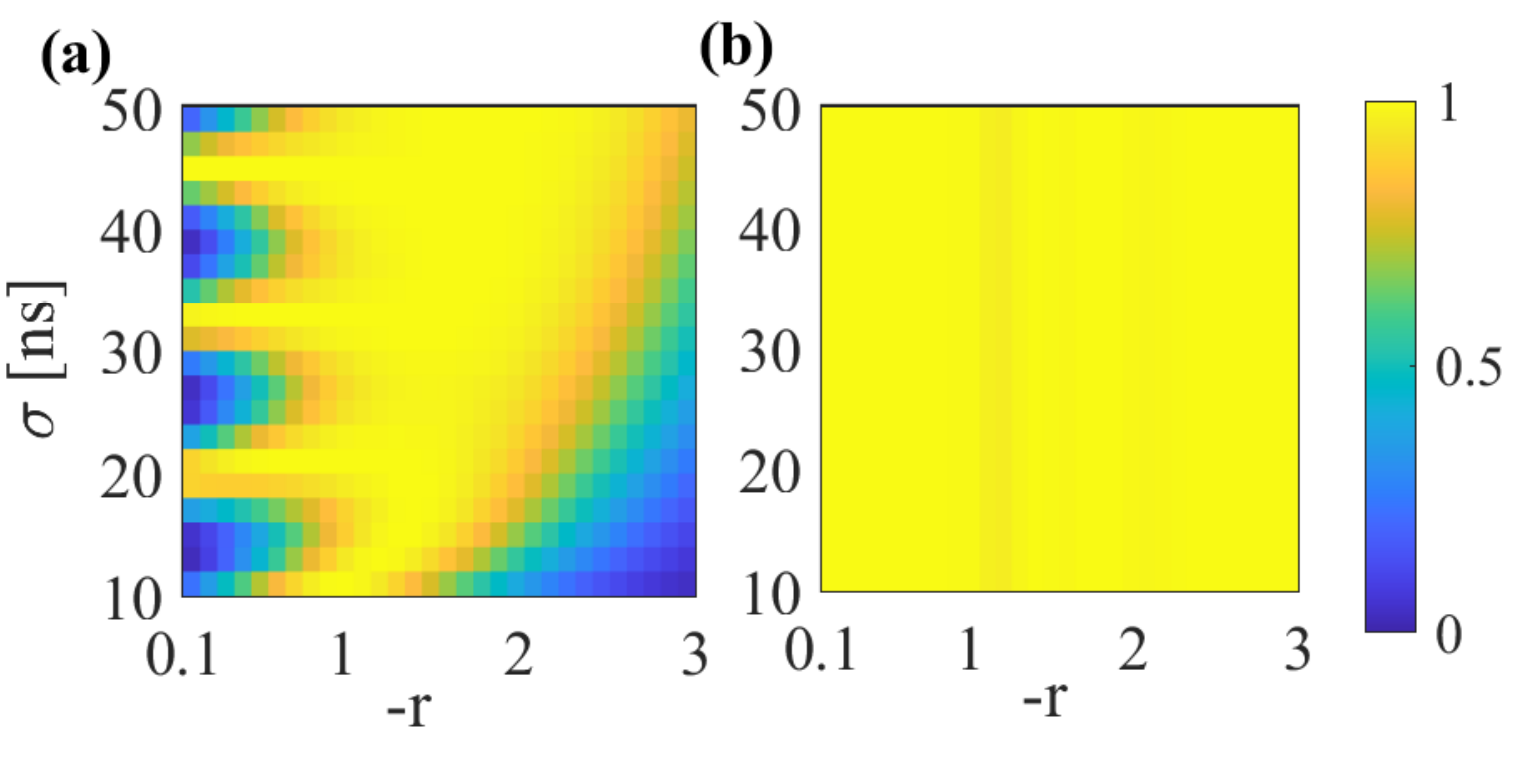}
 \caption{Population of the second excited state ($p_2$) is plotted at time $t=t_f$ in a STIRAP protocol, with initial state $|0\rangle$. Results from (a) STIRAP with $\Omega=45$ MHz and $n_t=3$ and (b) well-tailored STIRAP as per Eqs.~(\ref{eq-nt-r}) and (\ref{adia7r}) are shown.  }
 \label{Fig-comparison}
\end{figure}

A demonstration of the improvement achieved by employing the conditions in Eqs.~(\ref{eq-nt-r}) and (\ref{adia7r})
is shown in Fig.~\ref{Fig-comparison}, where population $p_2$ of the second excited state ($|2\rangle$) at the final time $t=t_f$ is plotted as a function of $\sigma$ and $-r$, with $\alpha=1$ and $\epsilon=0.05$.  Fig.~\ref{Fig-comparison}(a) shows the $p_2$ with fixed $n_t=3$  and $\Omega=45$ MHz. Thus, any arbitrary point on the $p_2$-map in Fig.~\ref{Fig-comparison}(a) satisfies $\sigma \Omega > \sqrt{\pi}/4$. The simulation of a perfectly tailored STIRAP utilizing the conditions in Eqs.~(\ref{eq-nt-r}) and (\ref{adia7r}) is shown in Fig.~\ref{Fig-comparison}(b). The resultant population profile demonstrates a well-tailored STIRAP protocol for the desired population transfer. 
Fig.~\ref{Fig-comparison} (b) shows clear
improvement relative to results shown in Fig.~\ref{Fig-comparison} (a).

For a  practical implementation, the total time-cost and pulse-power evaluation are also important. The corresponding maps of the total pulse duration and maximum pulse-amplitude in the same ranges of $\sigma$ and $-r$ are shown in the Appendix in  Fig.~7. 
The wide range of resultant $T$ and $\Omega$ values provide flexibility to the protocol. Larger values of $\epsilon$ further lead to significant reduction in the time-cost. In turn, a choice of slightly larger $\sigma$ can significantly reduce the amplitude $\Omega$.

Alternatively, when evaluating the experimental feasibility various parameters can be constrained and a perfect population transfer can be designed with the help of these interweaved parametric equations and graphs.

For a desired value of $\epsilon$, and with the help of Fig.~\ref{fig_r_vs_nt}, Eqs.~(\ref{eq-nt})or~(\ref{eq-T}), and Eq.~(\ref{adia7}),
one can easily obtain an experimentally feasible set of parameters $r$, $\sigma$, $n_t$, and $\Omega$ that 
leads to a perfect STIRAP. 
It is noteworthy that the efficacy of this perfect STIRAP protocol does not rigidly rely on the calculated parameters. In fact, parameters such as $n_t$, $\Omega$ can be considered as the respective lower bounds to achieve population transfer with infidelity $\epsilon^2$. Larger values of these parameters will only make the transfer more efficient. This makes the protocol robust against the experimental imperfections. 

\section{Discussion and conclusions}
We presented a well-tailored STIRAP protocol that leads to a perfect population transfer $|0\rangle \rightarrow -|2\rangle$ alongside with flexibility in the choice of parameters. For a given $\epsilon$, a combined choice of parameters $n_t$ and $r$  already determines the final population to be transferred. Furthermore, the choice of $\sigma$ determines the total pulse duration ($T$) and the corresponding calculated value of the amplitude $\Omega$ is responsible for the pulse power. A trade off between $\sigma$ and $\Omega$ values can be settled by evaluating the  experimental feasibility. We also discussed the relatively general situation, where the Gaussian drives can have unequal maximum amplitudes controlled by the parameter $\alpha=\Omega_{01}^{0}/\Omega_{12}^{0}$. The results for $\alpha \neq 1$ are discussed in the main text, while a detailed analysis is presented in the appendices.  The  analysis presented here relies on a simple set of calculations and observations, however the end results are 
non-trivial. In conclusion, our calculations for the STIRAP drives lead to a perfect population transfer within the reach of experimentally feasible scenario and without the help of any additional shortcuts to the adiabaticity.

\begin{acknowledgments} 
We acknowledge financial support from the Academy of Finland under the Finnish Center of Excellence in Quantum Technology QTF (projects 312296, 336810) and from Grant No. FQXi-IAF19-06 (“Exploring the fundamental limits set by thermodynamics in the quantum regime”) of the Foundational Questions Institute Fund (FQXi), a donor advised fund of the Silicon Valley Community Foundation.
\end{acknowledgments}

\appendix

\section{Practical feasibility}
 
 \begin{figure}[ht]
 \centering
 \includegraphics[width=8cm,keepaspectratio=true]{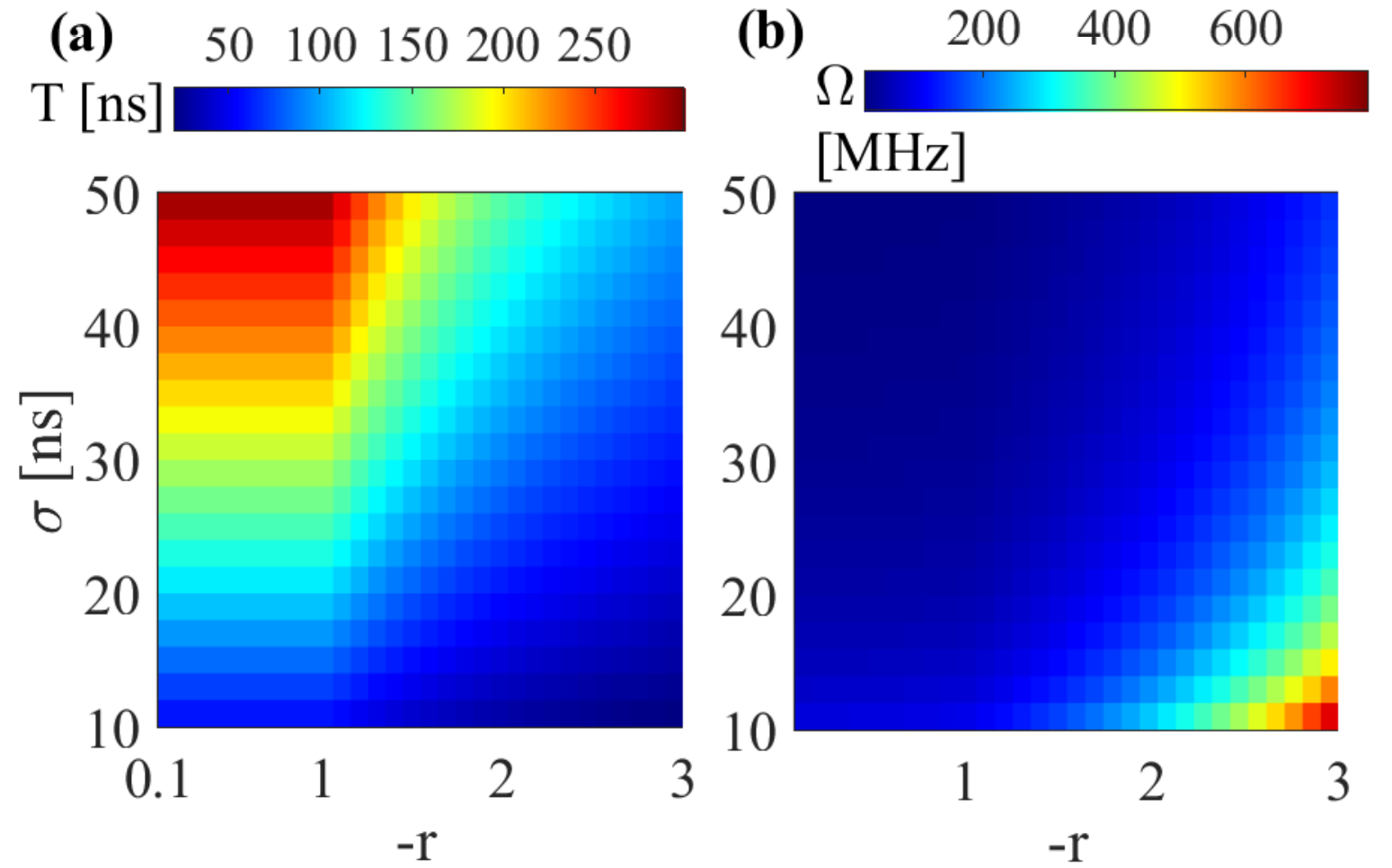}
\label{Fig-app2}
\caption{Variation of (a) total pulse duration (T) and (b) maximum pulse-amplitude ($\Omega$) as a function of $\sigma$ and $-r$.}
 \end{figure}
 
 The cost in terms of the total pulse duration and maximum pulse-amplitude is shown in Fig.~7 
 as surface maps for a wide range of $\sigma$ and $-r$, consistent with Fig.~\ref{Fig-comparison}(b) of the main text.

 
\section{Gaussians with different amplitudes \label{app-omega01neqomega12}}
Consider the pair of Gaussians together leading to a  STIRAP pulse sequence 
for $\Omega_{01}^{0} \neq \Omega_{12}^{0}$ and $\sigma_1=\sigma_2=\sigma$, 
\begin{eqnarray}
 \Omega_{01}(t) &=& \Omega_{01}^{0} e^{-t^2/2\sigma^2}, \nonumber \\
 \Omega_{12}(t) &=& \Omega_{12}^{0} e^{-(t-t_s)^2/2\sigma^2}.
\end{eqnarray}
The mixing angle is given by
\begin{eqnarray}
 \tan \Theta(t) &=& \frac{\Omega_{01}(t)}{\Omega_{12}(t)}=\frac{\Omega_{01}^{0}}{\Omega_{12}^{0}} e^{-rt/\sigma}e^{r^2/2}, \\ 
 \textrm{or} \quad \frac{rt}{\sigma} &=& \frac{r^2}{2} - \ln \left[ \tan \Theta(t) \right] + \ln \left[ \Omega_{01}^{0}/\Omega_{12}^{0} \right]. \label{eq-theta2_app}
\end{eqnarray}
where, $r=t_s/\sigma$. For a finite time operation, this sequence of Gaussians is truncated at 
the optimal time that provides the complete transfer of population.
The initial and final time points of the sequence are obtained by truncating the sequence 
from left at $t=t_i=-n_i \sigma +t_s$ and from right at $t=t_f=n_f \sigma $, such that the 
total operation time is $T=\sigma(n_i+n_f)-t_s$. Interestingly, we come across the same mixing angle ($\Theta$) in the structure
of the dark state $\vert \psi_D \rangle$, which is an eigenstate of the instantaneous
Hamiltonian with eigenvalue $0$. We assume our three-level quantum system
in the dark state with close proximity to the ground state ($\vert 0 \rangle$),
\begin{equation}
  \vert \psi_i \rangle 
 = \left( \begin{array}{c} \sqrt{1-\epsilon^2} \\ 0 \\ -\epsilon \end{array} \right), 
 \end{equation}
such that $\epsilon \longrightarrow 0$.
We know that during the evolution under the STIRAP Hamiltonian, our system is ideally in the dark state at 
all times, such that the final state is
\begin{equation}
  \vert \psi_f \rangle  = \left( \begin{array}{c} \epsilon  \\ 0 \\ -\sqrt{1-\epsilon^2} \end{array} \right). 
 \end{equation}
 At the final time point, $t=t_f=n_f \sigma$, and from Eqs.~(\ref{eq-si}),~(\ref{eq-sf}), and~(\ref{eq-theta2_app}), 
 one may easily arrive at
  \begin{equation}
  n_f = \frac{1}{r} \ln \frac{\epsilon}{\sqrt{1-\epsilon^2}} +\frac{r}{2}  +\frac{1}{r} \ln \frac{\Omega_{01}^{0}}{\Omega_{12}^{0}} \label{eq-ntf}.
 \end{equation} 
 Similarly, at  $t=t_i=-n_i \sigma+t_s$
   \begin{equation}
  n_i = \frac{1}{r} \ln \frac{\epsilon}{\sqrt{1-\epsilon^2}} +\frac{r}{2}  -\frac{1}{r} \ln \frac{\Omega_{01}^{0}}{\Omega_{12}^{0}} \label{eq-nti},
 \end{equation} 
 and the total time,
 \begin{equation}
  T=(n_i+n_f)\sigma-t_s=(n_i+n_f-r)\sigma=\frac{2\sigma}{r} \ln \frac{\epsilon}{\sqrt{1-\epsilon^2}}.
 \end{equation}
 Consistent with parameter Set 2, the above equations are valid for $|r|\geq 1$. For $|r|<1$ we have,

   \begin{eqnarray}
  n_f &=& -\ln \frac{\epsilon}{\sqrt{1-\epsilon^2}} +\frac{r}{2}  +\frac{1}{r} \ln \frac{\Omega_{01}^{0}}{\Omega_{12}^{0}} \label{eq-ntf}.
\\
  n_i &=& -\ln \frac{\epsilon}{\sqrt{1-\epsilon^2}} +\frac{r}{2}  -\frac{1}{r} \ln \frac{\Omega_{01}^{0}}{\Omega_{12}^{0}} \label{eq-nti},
 \\
  T &=& (n_i+n_f)\sigma-t_s=(n_i+n_f-r)\sigma \nonumber \\ 
    &=& 2\sigma \ln \frac{\epsilon}{\sqrt{1-\epsilon^2}}. 
 \end{eqnarray}

When $\Omega_{01}^{0} \neq \Omega_{12}^{0}$, truncation from the left and right extremes correspond to 
slightly different values of $n_i$ and $n_f$ to obtain the final state with infidelity  $\epsilon^2$.
This is due to the same amount of fractional decrease in the amplitudes of the pulses expected 
at the initial and final time points. 
	


%

\end{document}